# Local probe of the interlayer coupling strength of few-layers SnSe by contact-resonance atomic force microscopy

Zhiyue Zheng[1#], Yuhao Pan[2#], Tengfei Pei[3], Rui Xu[2], Kunqi Xu[4], Le Lei[2], Sabir Hussain[5,6], Xiaojun Liu[1], Lihong Bao[3], Hong-Jun Gao[3], Wei Ji[2]*, Zhihai Cheng[2]*

[1]*State Key Laboratory of Digital Manufacturing Equipment and Technology, School of Mechanical Science and Engineering, Huazhong University of Science and Technology, Wuhan, 430074, China*

[2]*Beijing Key Laboratory of Optoelectronic Functional Materials & Micro-nano Devices, Department of Physics, Renmin University of China, Beijing 100872, China*

[3]*Beijing National Laboratory for Condensed Matter Physics, Institute of Physics, Chinese Academy of Sciences, Beijing 100190, China*

[4]*Key Laboratory of Inorganic Functional Materials and Devices, Shanghai Institute of Ceramics, Chinese Academy of Sciences, Shanghai 200050, China*

[5]*CAS Key Laboratory of Standardization and Measurement for Nanotechnology, CAS Center for Excellence in Nanoscience, National Center for Nanoscience and Technology, Beijing 100190, China*

[6]*University of Chinese Academy of Sciences, Beijing 100049, China*

\# These authors contributed equally to this work.

\* Corresponding authors: zhihaicheng@ruc.edu.cn wji@ruc.ed.u.cn

**Abstract:** The interlayer bonding in two-dimensional (2D) materials is particularly important because it is not only related to their physical and chemical stability but also affects their mechanical, thermal, electronic, optical, and other properties. To address this issue, we report the direct characterization of the interlayer bonding in 2D SnSe using contact-resonance atomic force microscopy (CR-AFM) in this study. Site-specific CR spectroscopy and CR force spectroscopy measurements are performed on both SnSe and its supporting SiO$_2$/Si substrate comparatively. Based on the cantilever and contact mechanic models, the contact stiffness and vertical Young's modulus are evaluated in comparison with SiO$_2$/Si as a reference material. The interlayer bonding of SnSe is further analyzed in combination with the semi-analytical model and density functional theory calculations. The direct characterization of interlayer interactions using this non-destructive methodology of CR-AFM would facilitate a better understanding of the physical and chemical properties of 2D layered materials, specifically for interlayer intercalation and vertical heterostructures.

# 1 Introduction

The family of two-dimensional (2D) layered materials has been expanded appreciably since 2004, and a worldwide scientific and technological effort has been made to understand and control their properties owing to their potential technological applications [1-7]. One of the main characteristics of 2D layered materials is the strong in-plane bonds of the atomic layer and the relatively weak interactions between the atomic layers (i.e., interlayer interaction). The in-plane elasticity and strength have been widely investigated by atomic force microscopy (AFM) indentation experiments on suspended 2D films. However, little attention has been paid to the characterization of the interlayer interactions of 2D layered materials. Nevertheless, the interlayer interaction [8-12] is particularly important because it corresponds to not only physical and chemical stability but also mechanical, thermal, electronic, and optical properties of 2D layered materials. Lack of information on the interlayer interactions makes it difficult to study vertical Van der Waals (vdW) heterostructures that are developed by stacking different 2D atomic layers with their respective metallic, semiconducting, or insulating properties. It could be anticipated that in addition to normal vdW interactions, charge transfer and electrostatic dipole interactions between the neighboring layers could contribute to the unnatural interlayer bonding within vertical heterogonous structures. In addition, the interlayer intercalation and contamination further complicate the interlayer interactions of 2D materials or vdW heterostructures.

In general, it is necessary to develop a reliable measurement method, which can probe the interlayer interactions of 2D materials in a relatively direct and simple manner. Currently, several nanomechanical AFM techniques exist or are being developed, e.g., tip indentation (force-distance spectroscopy) [13,14], force modulation [15-18], and contact-resonance AFM (CR-AFM) [19-22]. The AFM tip indentation method has been widely used to study in-plane Young's modulus of 2D layered materials. Recently, Yang Gao et al. [23] have introduced a new modulated nano-indentation AFM method to study the vertical elasticity of few-layer-thick graphene and graphene oxide films, which is a force modulation method applied in force spectroscopy measurements. However, it is time-consuming and laborious to conduct the aforementioned measurements accurately or calibrate the cantilever spring constant, optical sensitivity, and tip radius, owing to unexpected changes of the tip condition during the experiments, which detracts from the simplicity of the original idea.

The newer CR-AFM technique has better advantages for the nanomechanical characterization of layered materials compared to the existing techniques. It is based on a comparative method that is performed on a test sample and a reference sample with known modulus. This comparative method avoids the complicated calibration of tip parameters as described above, which makes the measurements highly efficient and direct. In addition, it is achieved by introducing a slight vertical modulation to either the cantilever base (called ultrasonic atomic force microscopy (UFAM)) [24,25] or the sample (called atomic force acoustic microscopy (AFAM)) [26,27] as the tip is in contact with the sample, wherein the values of the cantilever CR frequency (CRF) and quality factor ($Q$) change in response to the viscoelastic properties of the sample. The CRF is significantly larger than the corresponding eigenfrequency; the typical Q factor in air for CR-AFM is about 10-100, which implies that a weak oscillation signal can be amplified by a factor of 10-100. Therefore, CR-AFM can quantitatively characterize the viscoelastic response of materials with high sensitivity and a large dynamic range of the detection because of the high-frequency operation and the resonance enhancement effect [22,28]. The detection limit of AFM used herein is approximately 10 pm; therefore, the excitation amplitude of the tip–sample contact can reduce to the picometer level, which is considerably smaller than the interlayer distance of 2D layered materials. Here, we report the direct characterization of the interlayer bonding in 2D SnSe based on the AFAM apparatus. Site-specific CR spectroscopy and CR force-spectroscopy measurements are performed on both SnSe and its supporting $SiO_2$/Si substrate comparatively. Using the cantilever and contact mechanic models, the contact stiffness and vertical Young's modulus are evaluated in comparison with $SiO_2$/Si as a reference material. The stronger interlayer bonding of SnSe was further analyzed in combination with the semi-analytical model and density functional theory (DFT) calculations.

## 2 Principle and quantitative analysis of CR-AFM

### 2.1 Principle and model of CR-AFM

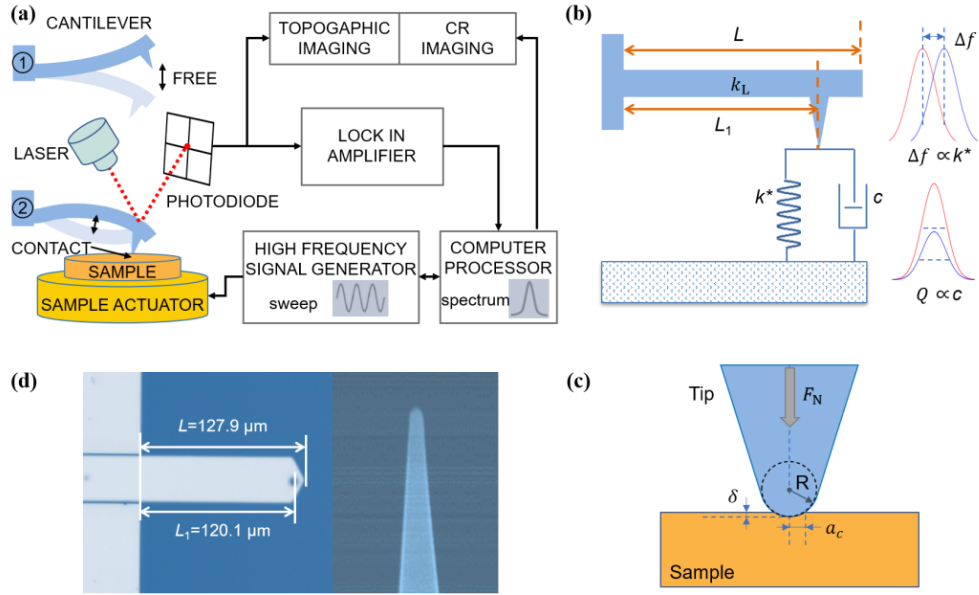

**Fig.1 Schematic diagram and mechanical model of CR-AFM and related definitions and characterization of the cantilever used in the experiments for quantitative analysis.** (a) Schematic diagram of the CR-AFM experimental setup. (b) Tip–sample contact is modeled as a Kelvin–Voigt mechanical equivalent, where the spring and the dashpot represent the contact stiffness (elastic modulus) and viscoelasticity (dissipation), respectively. (c) Close-up view of the tip–sample contact and definitions for quantitative characterization. (d) SEM characterization of the AFM cantilever and tip.

Fig. 1 shows the schematic diagram and mechanical model of CR-AFM and the related definitions and characterization for quantitative analysis. CR-AFM is based on the typical contact mode imaging method, and the key is the vertical modulation with a slight amplitude introduced to the sample, as illustrated in Fig. 1(a). The modulation is operated at high frequency and hence does not affect the contact-feedback imaging; however, the modulation can be coupled to the cantilever deflection that can be extracted for detection. The high frequency signal is continuously changing with the sample mechanical properties as the tip scans the sample in contact mode, which is detected by a lock-in amplifier and referred to the CRF and CR amplitude (CRA). Fig. 1(b) depicts the mechanical model of the tip–sample dynamic contact in CR-AFM. As the tip location on the cantilever directly affects the final measurement result in CR-AFM, the cantilever is modeled as a distributed mass with spring constant $k_L$ rather than the point-mass approximation. The tip location is considered as a relative position $\gamma = L_1/L$ (with $0 \leq \gamma \leq 1$) from the clamped end. The tip–sample interaction is considered to be normal elastic and dissipative (damping) forces, which are modeled as a spring and a dashpot in parallel (i.e., Kelvin–Voigt model). The variables $k^*$ and $c$ indicate the contact stiffness and damping, respectively. The variations in the tip–sample stiffness and damping can be reflected in the CRF and $Q$ of the coupled system, which can be converted to the sample elastic and loss moduli, respectively, with right contact mechanic models. Higher contact stiffness indicates higher CRF, and larger viscosity (dissipation) of the tip–sample contact indicates lower $Q$.

### 2.2 Modulus quantification in CR-AFM

In CR-AFM, the primary contact mechanical quantity of interest is the contact stiffness $k^*$, which characterizes the elastic interaction between the tip and the sample. Once $k^*$ is determined, the appropriate contact mechanical model is applied and thus the elastic modulus can be obtained. Therefore, the quantification of the sample modulus mainly depends on two mechanic models: cantilever dynamic model (Fig. 1(b)) and tip–sample contact mechanical model (Fig. 1(c)).

Consideration of the cantilever dynamic model is primarily based on the Euler–Bernoulli beam theory. The contact stiffness $k^*$ is calculated in terms of the cantilever stiffness, relative tip position, and wavenumber of the resonance [29], as described as Eqs. 1–4.

The relative tip position $\gamma$ is characterized by SEM, as shown in Fig. 1(d),

$$\gamma = \frac{L_1}{L} = \frac{120.1}{127.9} = 0.939 \tag{1}$$

The parameter $A_n$ related to the wavenumber $x_n L$ of the cantilever resonance in free space is defined by

$$A_n^2 = \frac{(x_n L)^2}{f_n^0} \tag{2}$$

The wavenumber is denoted as $y_n L$ under the tip–sample coupling condition:

$$y_n L = A_n \sqrt{f_n^{CR}} \tag{3}$$

Based on the above equations (Eqs. 1–3), the contact stiffness $k^*$ can be determined as shown in Eq. 4

$$k^* = \frac{2}{3} k_L (y_n L \gamma)^3 \frac{1 + \cos y_n L \cosh y_n L}{D} \tag{4}$$

Where

$$\begin{aligned}D = &[\sin y_n L(1-\gamma) \cosh y_n L(1-\gamma) - \cos y_n L(1-\gamma) \sinh y_n L(1-\gamma)][1 \\ &- \cos y_n L\gamma \cosh y_n L\gamma] \\ &- [\sin y_n L\gamma \cosh y_n L\gamma - \cos y_n L\gamma \sinh y_n L\gamma][1 \\ &+ \cos y_n L(1-\gamma) \cosh y_n L(1-\gamma)]\end{aligned}$$

The wavenumber $x_n L$ for the flexural mode $n$ is the solutions of the characteristic equation for cantilever vibration in free space $1 + \cos x_n L \cosh x_n L = 0$; the first two roots $x_1 L$ and $x_2 L$ are 1.8751 and 4.6941, respectively.

With the contact stiffness, then the sample modulus can be extracted quantitatively based on the accurate contact mechanic model. Typically, the contact mechanic models [29] include Hertzian (Hertz) contact model, Derjaguin-Muller-Toporov (DMT) model and Johnson–Kendall–Robert (JKR) model. While two frequently used models in CR-AFM are the Hertz contact and punch contact models, which depend on the shape of the tip (sphere or punch) [29]. The discussion here is restricted to the Hertz and DMT models due to the use of a spherical tip in all experiments. The main difference between the two models is the consideration of the adhesion force $F_{ad}$, regardless of CR-AFM. Fig. 1(c) depicts the Hertz contact of the sphere tip and the sample and defines the related quantities ($F_N$, $R$, $a_c$, and $\delta$). The normal contact stiffness of the tip–sample system can be expressed as shown in Eq. 5, which is related to the contact radius $a_c$ and reduced Young's modulus $E^*$.

$$k = 2a_c E^* \tag{5}$$

$$E^* = \left(\frac{1}{E_t} + \frac{1}{E_s}\right)^{-1} \tag{6}$$

$$a_c = \left(\frac{3 F_N R}{4 E^*}\right)^{1/3} \tag{7}$$

Where $E_s$ and $E_t$ are Young's moduli of the sample surface and AFM tip, respectively.

When the adhesion force $F_{ad}$ is considered in the measurement, the DMT model is accurate. The contact radius $a_{DMT}$ is modified as

$$a_{DMT} = \left(\frac{3(F_N - F_{ad})R}{4E^*}\right)^{1/3}. \tag{9}$$

Finally, the elastic modulus of the test sample can be determined from the experimental values of the contact stiffness by Eqs. 9–10.

$$E^* = E_{cal}^* \left(\frac{k_{test}^*}{k_{cal}^*}\right)^{-1} \tag{10}$$

The indentation can be deduced from the DMT model and is described as Eq. 11,

$$\delta = \frac{3(F_N - F_{ad})}{2k^*} \tag{11}$$

## 3  Results and discussion

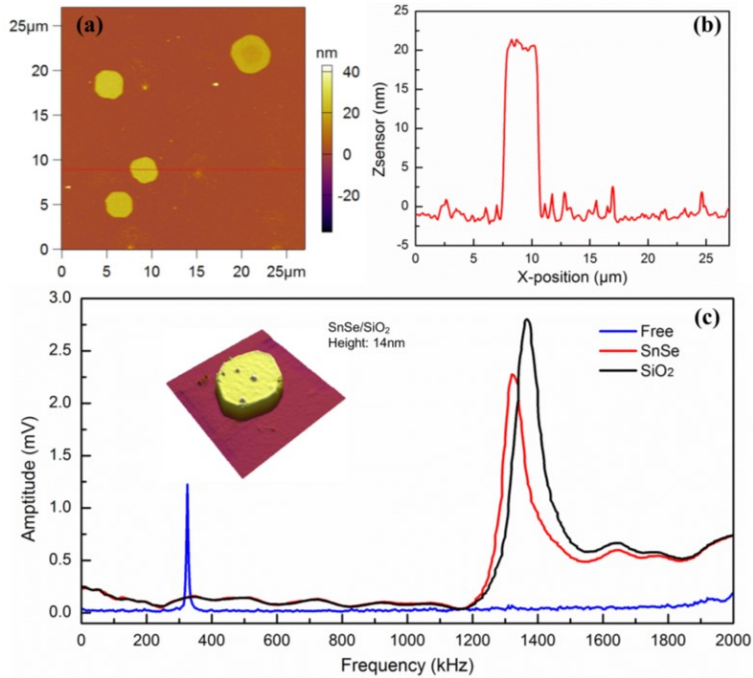

**Fig.2 CR-AFM spectroscopy measurements and imaging of SnSe on SiO$_2$/Si.** (a) Topography of SnSe on the SiO$_2$/Si wafer. (b) Corresponding line profile on the topography of SnSe on SiO$_2$/Si. (c) Site-specific CR-AFM spectroscopies obtained on SnSe and SiO$_2$/Si in a large frequency range. The frequency response of the free AFM probe is also supplied as a direct comparison with the CR-AFM frequency.

Fig. 2(a) shows a typical large-scale AFM topographic image of CVD-grown SnSe flakes on an SiO$_2$/Si substrate. A majority of the SnSe flakes show flat and uniform surfaces with typical heights of ~ 20 nm, as shown in Fig. 2(b). The inset of Fig. 2(c) shows a SnSe flake with a thickness of ~14 nm, which is thinner and flatter than most flakes and was thus used in our CR-AFM experiments. Site-specific CR spectroscopy measurements were performed at a fixed (stationary) position on the selected SnSe flake and the SiO$_2$/Si substrate. Specifically, the AFM probe was alternatively engaged on the SiO$_2$/Si substrate and the SnSe flake with the same loading force, and then the CR spectra were recorded, as shown in Fig. 2(c). The first flexural CRFs ($f_1^{CR}$) of both SnSe and SiO$_2$ are in the range of 1310–1370 kHz; however, the SnSe peak has a lower frequency compared to the SiO$_2$ signal, which indicates a lower modulus on the SnSe surface than that on the SiO$_2$/Si substrate. The first free flexural vibration of the cantilever ($f_1^0$) was also determined at ~327.4 kHz for the following calculations.

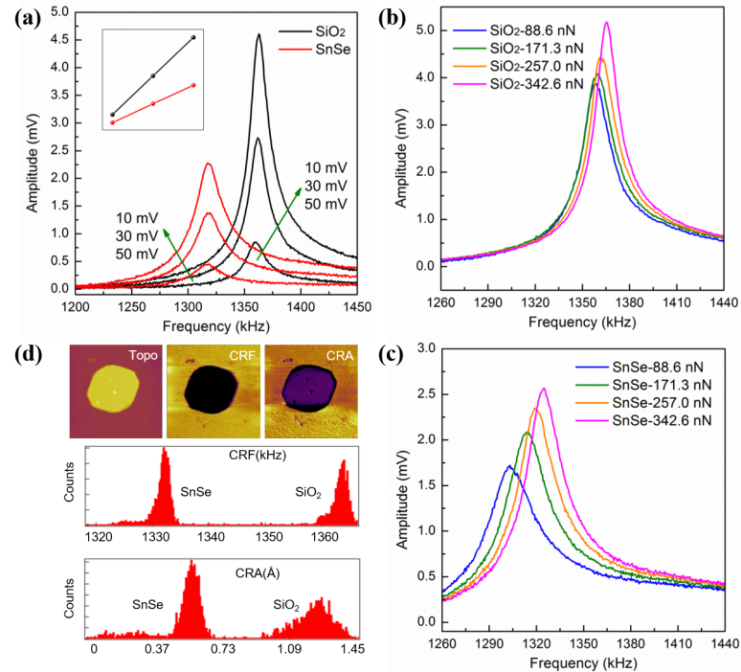

**Fig.3 Site-specific CR spectroscopy measurements and CR imaging.** (a) Site-specific CR spectroscopy at different excitations. (b) Site-specific CR spectroscopy at different loading force on the SiO$_2$/Si wafer. (c) Site-specific CR spectroscopy at different loading forces on SnSe. (d) Topography (Topo), CR images (CRF, CRA), and the corresponding histogram.

The detailed site-specific CR spectroscopy measurements were further performed at different loading forces and excitation voltages (applied to the sample actuator), as shown in Fig. 3. Fig. 3(a) demonstrates the detailed CR spectra at the same loading force (~88.6 nN) but different excitation voltages (10/30/50mV). The CRA on SnSe is lower than that on the $SiO_2$/Si substrate at the same loading force and excitations. The CRAs on the two samples increase almost linearly with the increasing driving amplitude, as shown in the inset of Fig. 3(a), whereas their CRFs remain almost unchanged. Fig. 3(b) and 3(c) demonstrates the detailed CR spectra at the same excitation (50mV) but different loading force (~88.6/171.3/257.0/342.6nN) on the $SiO_2$/Si substrate and SnSe, respectively. It is obvious that both CRFs and CRAs increase with the loading force, but the CRF on SnSe increases more significantly. The CR-AFM imaging measurements were further conducted, as shown in Fig. 3(d), which reveals a clear mechanical contrast between SnSe and the $SiO_2$/Si substrate. The corresponding histograms of the CRF and CRA show a narrower distribution on SnSe than on the $SiO_2$/Si substrate, indicating a better material quality of SnSe than that of the amorphous $SiO_2$/Si substrate. It is observed that the SnSe layer was gently scratched specifically at the edges of the SnSe island owing to the lateral shearing force during the CR-AFM imaging experiments.

Differing from the aforementioned CR-AFM spectroscopy measurements conducted under the specified loading force, the CR force spectroscopy measurements were further performed via the force-distance spectroscopy method, in which the CRF and CRA were also obtained simultaneously. It is noted that the experiments were conducted on $SiO_2$ and SnSe in an alternating way to ensure the consistent tip condition. This CR force spectroscopy allows us to obtain the relationship curves between different loading forces $F_N$ and CRF/CRA, as shown in Figs. 4(a) and 4(b). Both CRF and CRA increase nonlinearly, reach saturation with increasing the loading force and show a relatively different behavior on the $SiO_2$/Si substrate and SnSe. To determine the tip–sample contact mechanic properties, the $k^*$ as a function of $F_N$ on $SiO_2$ and SnSe was deduced from the measured free frequency and CRF as well as the geometric parameters of the AFM probe [30,31], as shown in Fig. 4(c). The results are consistent with the result of the hemispherical tip [32,33].

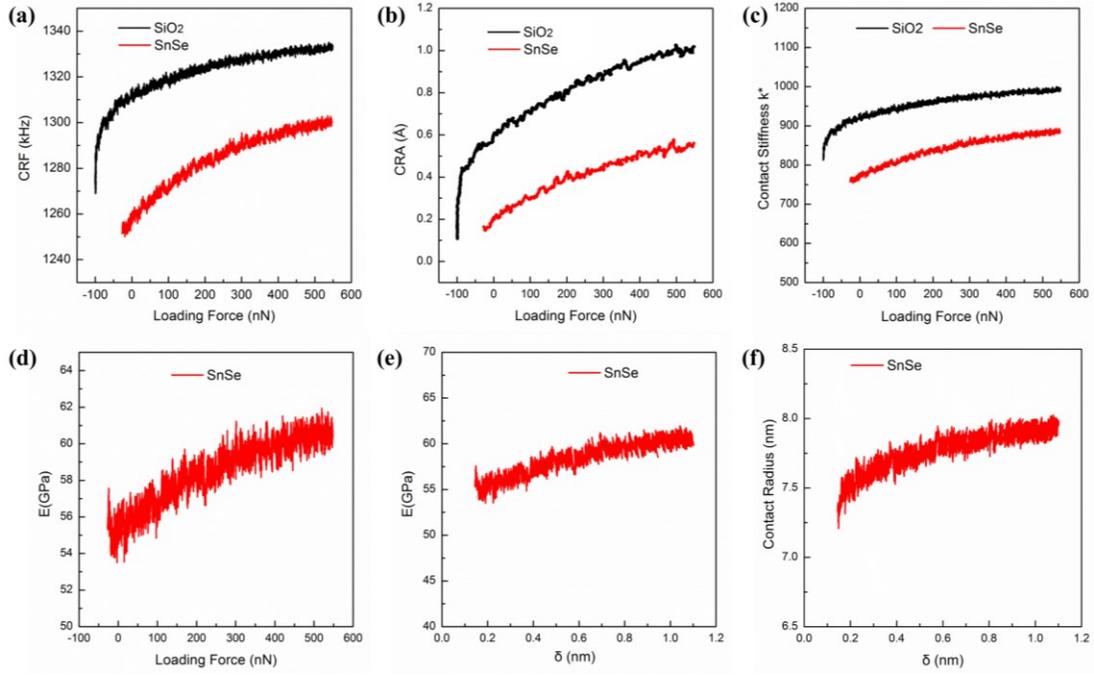

**Fig.4 CR-AFM force spectroscopy measurements.** (a) CRF as a function of the loading force $F_N$. (b) CRA as a function of the loading force $F_N$. (c) Contact stiffness as a function of the loading force $F_N$. (d) Vertical Young's modulus of layered SnSe as a function of the loading force. (E) Vertical Young's modulus of layered SnSe as a function of the indentation depth. (F) Contact radius as a function of the indentation depth.

After quantifying the contact stiffness $k^*$ and applying the appropriate contact mechanic model, the elastic moduli of the sample could be deduced. The prevailing model is the Hertzian (Hertz) model regardless of the adhesion force. However, the applied loading force is not greater than the adhesion force herein; hence, the use of the DMT model is more reasonable, which is an offset-Hertz model with the substitution of the applied loading force $F_N$ by the actual contact force $F_z = F_N - F_{ad}$. The adhesion force $F_{ad}$ was determined using force spectroscopy at about -100 nN for both SnSe and the SiO₂/Si substrate. Finally, the elastic modulus of the test sample (SnSe) can be determined from the contact stiffness. Young's modulus of the calibration sample (SiO₂/Si) and the tungsten carbide AFM tip are ~73 GPa [34] and ~714 GPa [35], respectively. Fig. 4(d) shows the curve of the elastic moduli of SnSe versus the loading force; the moduli are in the range of 55–61 GPa corresponding to the increasing loading force. The vertical elastic modulus and the contact radius on SnSe at different indentation depths can be obtained based on the DMT model, as shown in Figs. 4(e) and 4(f), which rise slightly with the indentation.

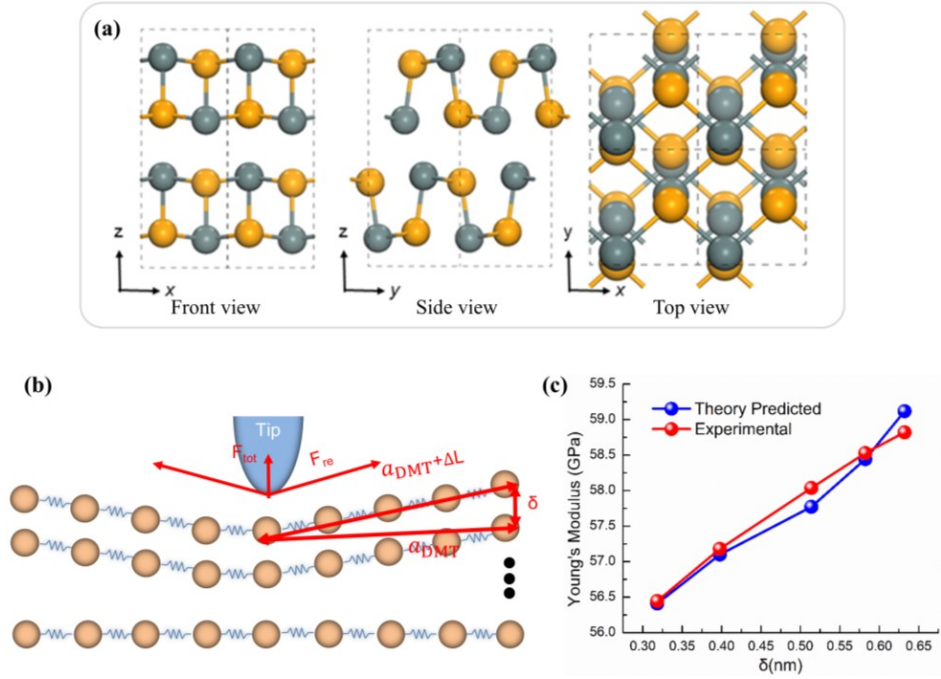

**Fig.5 Crystal structure and elastic modulus of SnSe predicted by the theoretical calculations.** (a) Front view (left), side view (middle), and top view (right) of the crystal structure of SnSe. (b) Semi-analytical model to estimate vertical Young's modulus of few-layer SnSe. $a_{DMT}$ and $\delta$ stand for the contact radius and the indentation depth, respectively. The length $\Delta L$ is the elongation of the SnSe lattice after indentation by the AFM tip. (c) Young's modulus as a function of the indentation depth by the theoretical computation and experiment.

Fig. 5 shows the crystal structure of SnSe, theoretical model, and vertical elastic modulus predicted by the DFT calculations. As shown in Fig. 5(a), SnSe shares a similar layered structure with black phosphorus (space group: Pnma, No. 62), but the Sn–Se bonds are buckled and have the AB stacking order. Fig. 5(b) illustrates an in-series spring model to estimate the vertical Young's modulus of layered SnSe. In this model, the interaction between the adjacent layers is simplified into ideal springs as the strain applied to layered SnSe is kept in a linear elastic region (<5%). In addition, the interactions between non-adjacent layers are ignored because their strengths are significantly weaker than those of the adjacent-layer interactions (<1/10). The Methods section contains more details of this model. According to the experimental measurement of the indentation depth ($\delta$) and contact radius ($a_{DMT}$), the AFM tip gives a bend to the surface of few-layer SnSe. The vertical component of the elastic restoring force ($F_{re}$) provides an auxiliary external force to the AFM tip and thus the estimated vertical Young's modulus increases. Here, we use a semi-analytical model to calculate the changes in vertical Young's modulus. The indentation depth ($\delta$) and the contact radius ($a_{DMT}$) were derived using the DMT model, as shown in Fig. 4(f), and the elongation ratio ($\Delta L/a_{DMT}$) was calculated by the Pythagorean theorem. The $F_{re}$ was calculated using the DFT (see more details in the Supplementary Materials). The vertical component of the $F_{re}$ was added to correct theory-predicted Young's modulus. Five points ($\delta$, $a_{DMT}$) on the curve in Fig. 4(f) are chosen to estimate the values of the changes in vertical Young's modulus. Vertical Young's modulus increases with the indentation depth (the smaller indentation depth indicates the interaction between fewer layers), as shown in Fig. 5(c). Experimentally measured and theory-calculated vertical Young's moduli are consistent, which verifies the reliability of our model. Therefore, our method will be a good choice to estimate vertical Young's modulus of a large range of 2D materials in the future, including promising heterostructures and transition metal dichalcogenides.

## 4　Conclusion

In summary, CR techniques are an emerging class of dynamic contact AFM methods that are a state-of-the-art approach to quantitative nanomechanical measurements. One prominent advantage of CR imaging is that it can achieve the detection of nanomechanical information from extremely small material volumes. Therefore, this work is aimed at proposing a new experimental and theoretical framework to investigate the interlayer mechanical coupling in 2D materials and their vertical heterogeneous structures. The CR-AFM methods can also be further developed to investigate the interlayer interactions under various external stimuli, such as electrical and magnetic fields. For example, the electrostatic response of interlayer interactions can be readily investigated by CR-AFM measurements under the electrical filed applied by the conductive tip, which is specifically important for vertical heterostructures made by stacking 2D layers with different properties. Charge transfer and build-in electric fields in such heterostructures could be locally probed by conductive CR-AFM measurements. Therefore, with these measurement innovations, the CR-AFM techniques will be more widely applied to various 2D materials and their heterostructures.

**Acknowledgment:** This project is supported by the Ministry of Science and Technology (MOST) of China (No. 2016YFA0200700 and No. 2018YFE0202700), National Natural Science Foundation of China (NSFC) (No. 21622304, 61674045, 11604063, 11622437, and 11974422), Strategic Priority Research Program, Key Research Program of Frontier Sciences, and Instrument Developing Project of Chinese Academy of Sciences (CAS) (No. XDB30000000, QYZDB-SSW-SYS031, YZ201418). Z. H. Cheng was supported by Distinguished Technical Talents Project and Youth Innovation Promotion Association CAS, Fundamental Research Funds for the Central Universities, and Research Funds of Renmin University of China (No. 18XNLG01 and No. 19XNQ025). Calculations were performed at the Physics Lab of High-Performance Computing of Renmin University of China and Shanghai Supercomputer Center.

## Materials and Methods

**Experimental setups:**

All experiments were performed using a commercial AFM system (Asylum Research, MFP-3D Infinity) equipped with an external lock-in amplifier (Zurich Instruments, HF2LI). Fig. 1(a) shows the schematic of the used CR-AFM setup. The sample under investigation was strictly glued on a high-bandwidth AFM sample actuator (Asylum Research), which provided a clean, linear, and broadband excitation. The CR spectrum of the tip–sample contact system was obtained by sweeping frequency with an external HF2LI, and the corresponding resonance peak was determined and recorded.

**CR-AFM probes:**

Hemispherical cone-shaped AFM probes (HSC-60-125C40-MC) were used in the CR-AFM experiments, as shown in Fig. 1(d). They were purchased from *Nanoscience Instruments* and manufactured by Team Nanotec. According to the supplied information, the width and length of the AFM cantilever are ~35 μm and ~125 μm, and the nominal resonance frequency and the spring constant are ~300 kHz and ~40 N/m, respectively. The AFM tip is coated with tungsten carbide (Young's modulus: ~714 GPa [35]) and a cone-shaped hemisphere with a radius of ~60 nm. By the calibration, the first resonance frequency $f_1^0$ and the spring constant $k_L$ are characterized to be ~327.4 kHz and ~35.7 N/m, respectively.

**Sample actuator:**

The CR techniques have very strict requirements on the excitation source. The ideal actuator would have a flat (free of resonances) and linear (linear with drive) response and high bandwidth (up to ~2 MHz). The CR sample actuator used herein was developed by Asylum Research and delivered decent performance for the quantitative CR-AFM measurements.

**Sample preparation:**

SnSe nanosheets were synthesized on polydimethylsiloxane (PDMS) substrates using the chemical vapor deposition method. After the epitaxy of SnSe on PDMS, the SnSe samples were transferred on an $SiO_2$/Si substrate by attaching the PDMS substrates directly. The detailed procedures are described in the previous report [36].

**Theoretical calculation and model:**

DFT calculations were performed using the generalized gradient approximation for the exchange-correlation potential, the projector augmented wave method [37,38], and a plane-wave basis set, as implemented in the Vienna ab-initio simulation package [39]. Density functional perturbation theory was employed to calculate the vibrational properties and to obtain the force constant matrix of few-layer and bulk SnSe. The kinetic energy cut-off for the plane-wave basis set was set to 500 eV for structural relaxation and 700 eV for the vibrational property calculations. A k-mesh of 24×24×1 was adopted to sample the first Brillouin zone of the conventional unit cell of few-layer SnSe. In the geometric optimization and vibrational property calculations, vdW interactions were considered at the vdW-DF [40,41] level with the optB886b exchange functional (optB86b-vdW)[42,43], which shows the best description of the lattice constant of bulk SnSe in our tested functionals, including optB88-vdW and PBE. The shape and volume of each supercell were fully optimized, and all atoms in the supercell were allowed to relax until the residual force per atom was lower than $10^{-4} \text{eV\AA}^{-1}$. The performances of the three functionals, i.e., PBE, optB88-vdW, and optB86b-vdW, were examined by calculating the lattice constants of SnSe bulk. The optB86b-vdW functional yielded the closest values to the experimental result [44]; see details in Table S1 of the Supplementary Materials. Hence, the optB86b-vdW functional was used for calculating the elastic properties of SnSe.

# Supplementary Materials

## Local probe of the interlayer coupling strength of few-layers SnSe by contact-resonance atomic force microscopy

Zhiyue Zheng[1,#], Yuhao Pan[2,#], Tengfei Pei[3], Rui Xu[2], Kunqi Xu[4], Le Lei[2], Sabir Hussain[5,6], Xiaojun Liu[1], Lihong Bao[3], Hong-Jun Gao[3], Wei Ji[2]*, Zhihai Cheng[2]*

[1]The State Key Laboratory of Digital Manufacturing Equipment and Technology, School of Mechanical Science and Engineering, Huazhong University of Science and Technology, Wuhan, 430074, China

[2]Beijing Key Laboratory of Optoelectronic Functional Materials & Micro-nano Devices, Department of Physics, Renmin University of China, Beijing 100872, China

[3]Beijing National Laboratory for Condensed Matter Physics, Institute of Physics, Chinese Academy of Sciences, Beijing 100190, China

[4]Key Laboratory of Inorganic Functional Materials and Devices, Shanghai Institute of Ceramics, Chinese Academy of Sciences, Shanghai 200050, China

[5]CAS Key Laboratory of Standardization and Measurement for Nanotechnology, CAS Center for Excellence in Nanoscience, National Center for Nanoscience and Technology, Beijing 100190, China

[6]University of Chinese Academy of Sciences, Beijing 100049, China

The raw measured site-specific CR-force spectroscopy is shown as Fig. S1, where the zero-loading force is normalized to the free condition, the negative value refers to attraction and the positive value means repulsion. The higher CRF and CRA on $SiO_2$/Si than SnSe as above can be observed, which increase differently with the approaching of the tip to sample. Note that the CRA in the paper is calculated with the first optical sensitivity.

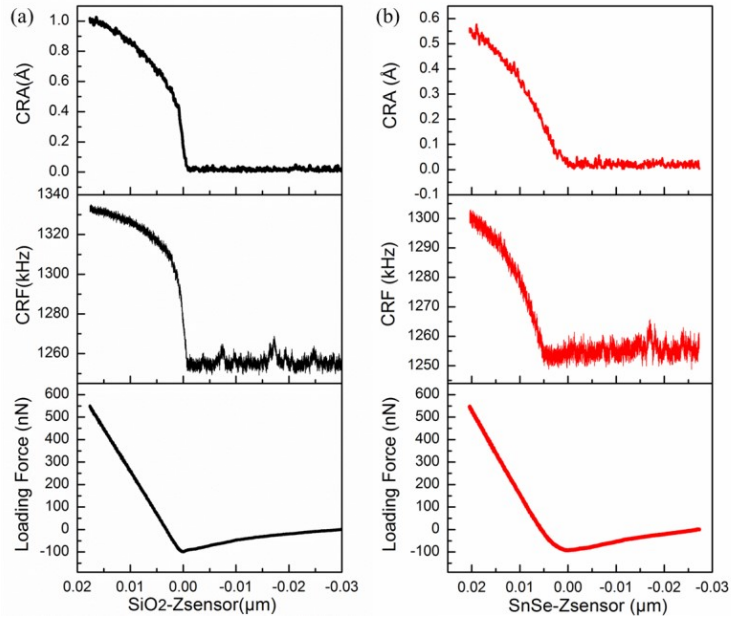

**Fig. S1 Raw data of Contact-Resonance Force Spectroscopy measurements on SnSe and its supporting substrate of SiO$_2$/Si.** (a) The changes of loading force, CRF and CRA with the approaching of the tip to SiO$_2$/Si. (b) The changes of loading force, CRF and CRA with the approaching of the tip to SnSe.

**Table** S1 Calculated lattice constant of our tested three functionals compared with experimental results. All three functionals overestimate the lattice constant along x direction. The functional optB86b performs best.

| Lattice Constant (Å) | a (x) | b (y) | c (z) |
|---|---|---|---|
| optB86b | 4.20 | 4.43 | 11.62 |
| optB88 | 4.21 | 4.53 | 11.78 |
| PBE | 4.21 | 4.56 | 11.79 |
| expt. | 4.14 | 4.44 | 11.70 |

**Calculation of force constant and Young's Modulus:**

**Projected inter-layer force constant.** In a rigid layer vibrational mode, the whole layer can be treated as one rigid body [1]. The projected inter-layer force constant, p-ILFC $k^i$ ($i$ stands for the projected direction, e.g. x, y or z) was constructed by summing inter-atomic force constants over all atoms from each of the two adjacent layers, as $k^i_{AB} = \sum_{a,b} D^i_{a,b}$ ($a \in$ [atoms in layer $A$], $b \in$ [atoms in layer $B$]). The matrix of inter-atomic force constants, essentially the same as the Hessian matrix of Born-Oppenheimer energy surface, is defined as the energetic response to a distortion of atomic geometry in DFPT [2]. It reads, $D_{a,b}^i = \frac{\partial^2 E_0(R)}{\partial R_a^i \partial R_b^i}$, where $R$ is the coordinate of each ion, $E_0(R)$ is the ground-state energy. Thus, the vertical p-ILFC defined as $\sum_{a,b} \frac{\partial^2 E_0(R)}{\partial R_a^i \partial R_b^i}$ approximately equals to the elastic constant defined as $\frac{\partial^2 E_0(R)}{\partial L_{ab}^2}$ ($L_{ab}$ represents the distance between a and b layer) in linearization range.

Young's modulus, defined as $E = \frac{\sigma}{\varepsilon}$ ($\sigma$ stands for the normal stress and $\varepsilon$ stands for the normal strain), were obtained in this work as follows:

$$E = \frac{\Delta F/A}{\Delta h/h_0} = \frac{h_0}{A} \cdot \frac{\partial F}{\partial h}$$

Therein, $F$ represents the external force, $A$ is the cross-sectional area, $h_0$ represents the effective layer thickness and $\Delta h$ is the thickness variations. In this work, $h_0$ was defined as $d^T + d^a$ ($d^T$ represents the thickness between the top layer and the bottom layer and $d^a$ represents average interlayer distance respectively). $\frac{\partial F}{\partial h}$ is the total vertical elastic constant of the system. An in-series-springs model was used to estimate the value of $\frac{\partial F}{\partial h}$. Each sublayer in SnSe was considered as an individual layer in this model. The elastic constant of each spring was given by p-ILFC as stated above. The total elastic constant $\frac{\partial F}{\partial h}$, therefore could be obtained according to spring series formula:

$$\frac{\partial F}{\partial h} = k_{tot} \approx \frac{1}{\sum_{i=1}^{2N-1} \frac{1}{k_{i,i+1}}} = \frac{1}{\sum_{i=1}^{2N-1} \frac{1}{\sum_{a,b} D_{a,b}^{vertical}}}$$

$a \in$ [atoms in the $i^{th}$ sublayer], $b \in$ [atoms in the $(i+1)^{th}$ sublayer B]

Here, $N$ is the number of SnSe layers. And $k_{i,i+1}$ represents the elastic constant between the $i^{th}$ and $(i+1)^{th}$ layers which equals to the vertical p-ILFC.

**Calculation of elastic restoring force.** The elastic restoring force ($F_{re}$) was calculated by multiplying the external pressure by the lateral surface area. The SnSe lattice was applied by a series of hydrostatic pressure from 0.0 to 3.0 kBar with a step of 0.5 kBar and fully relaxed to obtain the lattice changes. The pressure corresponding to the same value of elongation ratio with the experimental results could be obtained by interpolating the ratio-pressure curve.